# Morphodynamics of melting ice over turbulent warm water streams

Diego Perissutti[a], Cristian Marchioli[a] and Alfredo Soldati[b,a,*]

[a]*Polytechnic Department, University of Udine, Udine, 33100, Italy*
[b]*Institute of Fluid mechanics and Heat Transfer, TU Wien, Wien, 1060, Austria*

## ABSTRACT

We investigate the morphodynamics of an ice layer over a turbulent stream of warm water using numerical simulations. At low water speeds, characteristic streamwise undulations appear, which can be explained by the Reynolds analogy between heat and momentum transfer. As the water speed increases, these undulations combine with spanwise ripples of a much greater length scale. These ripples are generated by a melting mechanism controlled by the instability originating from the ice-water interactions, and, through a melting/freezing process, they evolve downstream with a migration velocity much slower than the turbulence characteristic velocity.

## 1. Introduction

The morphodynamics of basal melting and freezing of ice over warm deep-water currents is characterized by complicated interface patterns that are commonly observed under ice shelves (Rignot et al., 2013; Pritchard et al., 2012; Hirano et al., 2023). These patterns depend on heat and mass transfer but also on the water stream velocity (Gilpin et al., 1980; Bushuk et al., 2019). In particular, above a certain velocity threshold, the interface morphology exhibits features that can feedback on global melt rates and ice pack stability (Ian and Richard B., 2011; Alley et al., 2016). Precise appraisal of these phenomena has major direct implications on the quantification of global melt rates (Davis and Nicholls, 2019) and, in turn, on global climate predictions. Although the attention of the scientific community is high and the morphodynamics of ice-water interfaces has been widely studied (Washam et al., 2023; Claudin et al., 2017; Ashton and Kennedy, 1972), a comprehensive explanation of the physical mechanisms controlling the interface evolution is still elusive (Bushuk et al., 2019). At present, most ice melting models consist of empirical correlations (Holland and Jenkins, 1999) or heat transfer estimates valid for isothermal flat plates, as in the case of basal melting of icebergs (Cenedese and Straneo, 2023) or ice shelves (Dinniman et al., 2016; Goldberg et al., 2019). As a consequence, melt rate predictions can be inaccurate up to one order of magnitude (Jourdain et al., 2020; Davis et al., 2023; Nakayama et al., 2019) and cannot be reconciled with the large scatter of experimental data (Bushuk et al., 2019). Morphodynamic effects on local heat and mass transfer are crucially necessary to improve model accuracy and yet how this effect can be factored into existing parameterizations is still an open issue (Bushuk et al., 2019).

In the pioneering works by Ashton and Kennedy (1972), Hsu et al. (1979), the formation of spanwise wavy patterns, later reproduced in laboratory experiments (Gilpin et al., 1980; Bushuk et al., 2019), was observed. Pattern formation was attributed to the occurrence of a phase shift between the local heat transfer and the local ice thickness, and a positive growth rate of the interface deformation was predicted for phase shifts larger than $\pi/2$ (Ashton and Kennedy, 1972). It was also noted that turbulent mixing must be strong enough to trigger and sustain the shift. In the direct numerical simulations (DNS) by Couston et al. (2021), however, only streamwise-oriented canyons were found to form spontaneously, their spacing being compatible with that of near-wall turbulent velocity streaks. The emergence of these structures can be explained via the Reynolds analogy: Velocity fluctuations bring warm water to the ice-water interface and cold water away from it, favouring melting in regions of high shear stress and freezing in regions of low shear stress. Yet, the Reynolds analogy cannot explain the pattern formation mechanism discussed by Ashton and Kennedy (1972); Thorsness and Hanratty (1979), which was hypothesized to depend on anomalies in pressure and turbulent convection induced by the surface morphology. The causal relationship between these anomalies and the phase shift is still unclear, and no general closure model for predicting the shift as a function of the water turbulence properties is available.

*Corresponding author

In this work, we aim to reconcile the different morphodynamics just discussed and clarify the role played by turbulence in the formation (and possible coexistence) of the streamwise and spanwise interface patterns. In particular, we explore the physics that determine the extent of the phase shift as well as its dependence on the flow Reynolds number, not yet clarified in spite of the numerous numerical studies on ice melting (Yang et al., 2023a, 2024; Weady et al., 2022; Couston et al., 2021). To do so, inspired by Ashton and Kennedy (1972), we speculate that the change in pattern formation is associated to the existence of a critical threshold for the water stream velocity. We thus study how the features of the ice morphology depend on the flow conditions, by performing DNS at both sub-critical and super-critical water velocities.

## 2. Physical problem and Methodology

The physical problem investigated is the melting of an horizontal layer of ice under which a fully-developed turbulent shear flow is maintained. The flow domain is sketched in figure 1. To simulate a deep-water stream, a free-shear condition is imposed to the bottom boundary of the water layer, kept at bulk temperature $T_H$. The ice-water interface is kept at melting temperature $T_M$, whereas a no-slip condition is applied to the upper boundary of the ice layer, kept at bulk temperature $T_C$. To describe the evolution of temperature field $\mathcal{T} = (T - T_M)/\Delta T$, with $\Delta T = T_M - T_C = T_H - T_M$ and the velocity field $\mathbf{u}$, simulations are performed solving the continuity, Navier-Stokes and energy equations for incompressible water flow. The ice-water interface evolution and the ice volume fraction are computed using a phase field method (Hester et al., 2020; Yang et al., 2023b; Roccon et al., 2023; Soligo et al., 2021), which allows to capture the interface without introducing ad-hoc boundary conditions. These equations, in

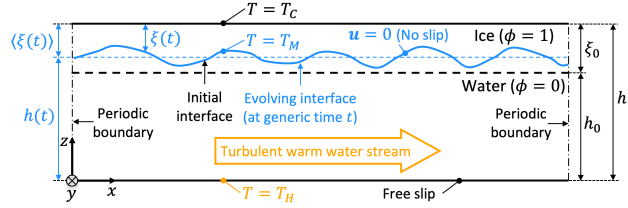

**Figure 1:** Schematic representation of the computational domain. The ice-water interface (black dashed line) is initially flat and positioned at $z = h_0$, with $h_0 = 0.75h$. As melting takes place, the interface deforms (blue line) and its position is defined by the local ice thickness $\xi(t)$.

dimensionless form, read as (Hester et al., 2020; Yang et al., 2023b):

$$
\begin{cases}
\nabla \cdot \mathbf{u} = 0 \\[2mm]
\dfrac{\partial \mathbf{u}}{\partial t} = -\nabla p + \dfrac{h_0}{h \text{Re}_{\tau,0}} \nabla^2 \mathbf{u} + \mathbf{S} \\[2mm]
\dfrac{\partial \phi}{\partial t} = \dfrac{5}{6} \dfrac{h_0}{hC\text{Re}_{\tau,0}\text{PrSt}} \nabla^2 \phi + S_\phi \\[2mm]
\dfrac{\partial \mathcal{T}}{\partial t} = \dfrac{h_0}{h\text{Re}_{\tau,0}\text{Pr}} \nabla^2 \mathcal{T} + \text{St} \dfrac{\partial \phi}{\partial t} + S_\mathcal{T}
\end{cases}
$$

where the non-linear terms $\mathbf{S}$, $S_\phi$ and $S_\mathcal{T}$ read as follows:

$$
\begin{cases}
\mathbf{S} = -\mathbf{u} \cdot \nabla \mathbf{u} - \dfrac{\phi^2}{\eta_s} \mathbf{u} \\[2mm]
S_\phi = -\dfrac{5}{6} \dfrac{h_0}{hC\epsilon^2 \text{Re}_{\tau,0}\text{PrSt}} \phi (1 - \phi)(1 - 2\phi + C\mathcal{T}) \\[2mm]
S_\mathcal{T} = -\mathbf{u} \cdot \nabla \mathcal{T}
\end{cases}
$$

The governing dimensionless numbers are the Stefan number, St, Prandtl number, Pr, and shear Reynolds number $\text{Re}_{\tau,0}$, based on the initial water layer height, $h_0 = 0.75h$ with $h$ the total height of the computational domain. A volume-penalization immersed boundary method, in which $\eta_s$ is the characteristic time scale, is used to account for the presence of the solid ice boundary in the Navier-Stokes equations. The phase field method is employed to obtain the evolution of the phase indicator $\phi$, which physically represents the local volume fraction of the ice phase (hence, $\phi = 1$ inside the ice layer and $\phi = 0$ inside the water layer, the position of the ice-water interface being located at $\phi = 0.5$). The thickness of the transition layer between ice and water is controlled by the phase-field parameter $\epsilon$. The characteristic time scale of the phase field depends on the constant $C$, which controls the dependence of the melting point on the interface curvature. Following (Yang et al., 2023b), the Gibbs-Thompson effects can be neglected for the present flow configuration and the value $C = 10$ can be chosen. The governing equations are advanced in time using an implicit scheme (Crank-Nicolson for the Navier-Stokes equations, implicit Euler for the other equations) for the linear diffusive terms and an explicit scheme (Adams-Bashforth) for the non-linear terms. At the initial time step, an explicit Euler method is employed for all the non-linear terms appearing in the equations, which are then solved in the spectral space by performing a discrete Fourier transform along the homogeneous flow directions and a discrete Chebyshev transform along the non-homogeneous flow direction orthogonal to the ice layer (Canuto et al., 2007). The computational domain is discretized using a three-dimensional Cartesian grid in which the axes $x$, $y$ and $z$ are aligned with the streamwise, spanwise and wall-normal directions, respectively. In line with Yang et al. (2023b), here we focus on the St = 0.1, Pr = 1 case. To examine both sub-critical and super-critical conditions, two different Reynolds numbers, $Re_{\tau,0} = 170$ and $636$, were chosen. These values were selected considering that the characteristic wavelength $\lambda_{x,cr}^+$ of unstable disturbances in our flow should be larger than 2100 in wall units, with the fastest growth rate being achieved at $\lambda_{x,cr}^+ \approx 3500$ (Hsu et al., 1979). The corresponding (estimated) critical Reynolds numbers are: $Re_\tau^{cr} = 251$ to have unstable modes, and $Re_\tau^{cr} = 418$ to have the fastest growing mode. Wall units are obtained using the fluid kinematic viscosity $\nu$ and the friction velocity $u_\tau = \sqrt{\tau_w/\rho}$, with $\tau_w$ the average shear stress at the ice-water interface and $\rho$ the water density.

## 3. Results

Figure 2 shows the morphological features of the ice-water interface at the end of the simulations. The interface morphology at low $Re_{\tau,0}$ (sub-critical case, panel a) is characterized only by the presence of streamwise-oriented structures (Couston et al., 2021), becoming more complex at high $Re_{\tau,0}$ (super-critical case, panel b), due to the emergence of prominent spanwise-oriented wavy patterns superposed to finer streamwise-oriented structures. These structures can be better appreciated in the windward portions of the interface. The streamwise structures are common to both flow regimes and indeed exhibit similar features when rescaled in wall units, as can be seen in figure 3. Panels a) and b) in this figure show the ice thickness maps at sub-critical and super-critical conditions, respectively. Panel c) shows the rescaled profiles of the ice-water interface, $\langle \xi^+ \rangle \xi'^+$, corresponding to a 200 wall units long, spanwise portion of the domain, indicated by sections A-A and B-B, respectively. The typical length scale of the wavy patterns is roughly equal to $\lambda_y^+ \sim 115$, a value that is of the same order as the characteristic size of near-wall turbulent streaks ($\lambda^+ \approx 100$) (Bernardini et al., 2014).

The spanwise patterns that characterize the super-critical regime are further examined in figure 4, which shows the ice thickness profile taken from a streamwise slice of the ice-water interface at high $Re_{\tau,0}$. The quasi-periodic behavior of the ice thickness can be traced back to the ripples observed by Ashton and Kennedy (1972); Gilpin et al. (1980) and extensively investigated by Hanratty (1981). Confirming the predictions formulated in these studies, we find that ripples grow in time but also shift downstream due to ice melting near the windward portion of the ice-water interface and water freezing in the leeward portion. The migration velocity $c$ of the ice ripples is computed from the spectrum of the cross-correlation between $\xi(t)$ and $\xi(t_f)$, which allows to measure the phase shift between the ice thickness at a generic time $t$ of the melting process and the thickness at the final time of the process in the simulation, $t_f$. For additional details, the reader is referred to Appendix A. Once the time derivative of the phase shift $\varphi$ is known, the migration velocity can be computed as follows:

$$c = \lambda_x^+ \frac{\partial\varphi}{\partial t^+},$$

where $\lambda_x^+$ is the characteristic wavelength of the ripples, in wall units. The migration velocity $c$ of the ripples is found to be significantly smaller than the velocity scale of near-wall turbulence, namely the friction velocity $u_\tau$: $c \approx 0.15u_\tau$. To quantify the difference between the sub-critical and super-critical ice patterns, we performed a spectral analysis of the streamwise and spanwise ice thickness profiles. The resulting spectra are shown in Figures 5a) and 5b), respectively. Both the amplitude and wavenumber in these panels are rescaled in wall units. Amplitude is additionally rescaled by multiplying $|\hat{\xi}^+|$ by the average ice thickness $\langle\xi^+\rangle$ to account for the characteristic response time of the ice interface, $\tau_i^+$, which depends on the strength of the diffusive transport through the ice layer and is thus proportional to $\langle\xi\rangle$. Note that, at high frequencies, the amplitude of the disturbances on the interface is inversely proportional to $\tau_i^+$, which is larger than the time duration of the thermal forcing (driven by turbulent fluctuations). Therefore, the ice does not have enough time to absorb all the energy content of the forcing and the amount of energy transferred mostly depends on the duration of the thermal fluctuations. From figure 5a), we observe that the sub-critical and super-critical spectra overlap in the medium-high wavenumber range, This provides further evidence that, in both regimes, the streamwise patterns are generated by the same turbulence-controlled mechanism and their features do not vary with the Reynolds number. The inset in figure 5a) shows the pre-multiplied spanwise spectra as a function of the wavelength $\lambda_y^+$. These spectra give an estimation of the spectral energy density associated with the ice-water interface, and ultimately to the surface area (which scales as $k_y^+ |\hat{\xi}_y|$ for sinusoidal profiles). The region of maximum energy density occurs around $\lambda_y^+ \approx 100$: This maximum is associated with the typical size of the near-wall turbulent coherent structures, and confirms the role of turbulence in determining the characteristic spanwise spacing observed in figure 3.

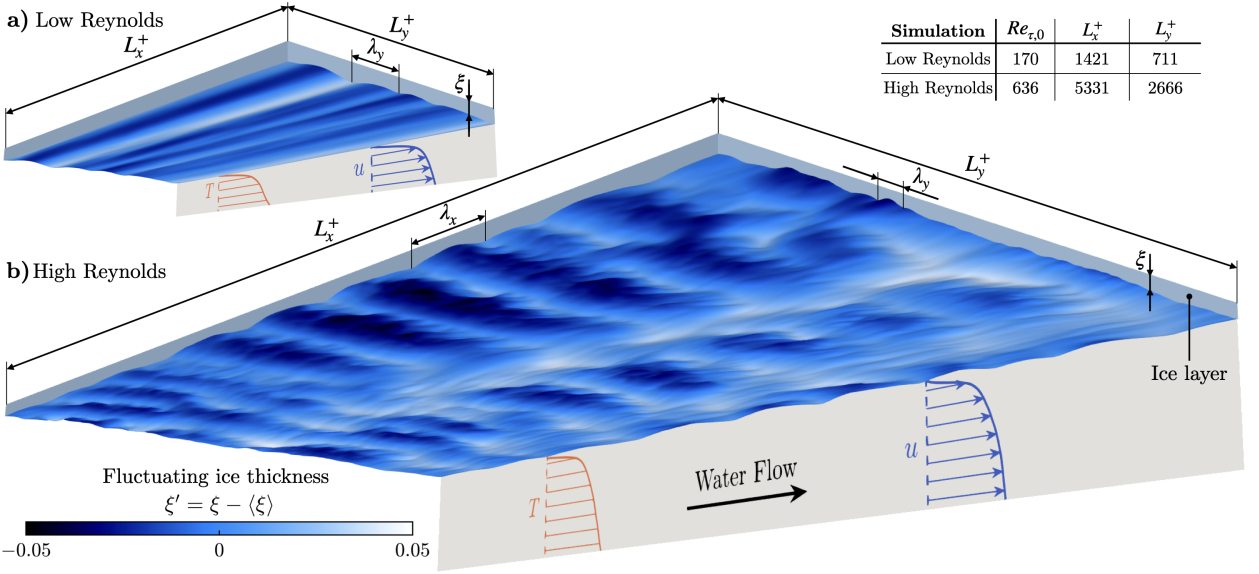

**Figure 2:** Morphology of melting ice over turbulent water flow in the sub-critical regime (panel a, $Re_{\tau,0} = 170$) and in the super-critical regime (panel b, $Re_{\tau,0} = 636$). The ice layer is displayed on top while water flows underneath. In both panels, light blue represents regions of higher-than-mean ice thickness, $\xi$, where the thickness fluctuation $\xi' = \xi - \langle\xi\rangle > 0$; dark blue represents regions of lower-than-mean thickness, where $\xi' < 0$. Note that the ice layer is comparatively thinner in panel b) due to stronger heat convection in the water. The computational domain size, given in dimensionless wall units in the table, as well as the wavelengths $\lambda_x$ and $\lambda_y$ of the surface patterns are also shown.

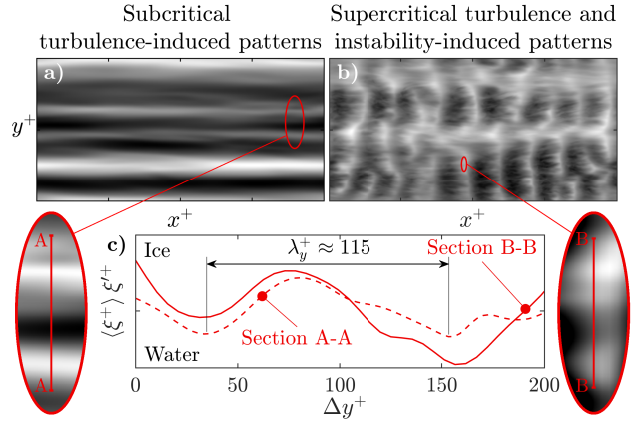

**Figure 3:** Ice thickness maps at low $\mathrm{Re}_{\tau,0}$ (panel a) and high $\mathrm{Re}_{\tau,0}$ (panel b). Thick ice regions are color-coded in white, thin ice regions are color-coded in black. The red segments labelled A-A and B-B in the insets have equal length in wall units and correspond to the spanwise portion of the interface from which the profiles shown in panel c) are taken.

Examining the streamwise spectra in figure 5b), a peak at $k_x^+ \simeq 1.5 \times 10^{-3}$ is visible at high $\mathrm{Re}_{\tau,0}$. At low $\mathrm{Re}_{\tau,0}$, instead, no peak is present because no prominent spanwise structure is formed. The peak value corresponds to $k_x L_x = 8$, implying that there are 8 crests across the domain length $L_x$, indicating that the peak can be directly associated to the spanwise patterns caused by the morphodynamic instability. The inset in figure 5b) shows that this peak, and hence the amplitude of the ripples, grows over time: The peak is not present in the early stages of ice melting, but forms gradually and reaches its maximum value at the end of the simulations. A necessary condition for the instability to occur is that the heat transfer must be shifted at least by a phase angle $\varphi_{q-\xi} = \pi/2$ with respect to the interface position (Gilpin et al., 1980). The mechanism that determines this phase shift is not fully understood but several authors (Ashton and Kennedy, 1972; Bushuk et al., 2019; Hanratty, 1981) have suggested that it could be due to turbulent fluctuations. What we observe is that, in super-critical conditions, regions of high turbulent kinetic energy form downstream of the crests; in these regions turbulent mixing is enhanced and, in turn, both the local effective momentum and thermal diffusivity are increased. As a result, the regions of maximum heat flux and maximum wall shear stress are shifted upstream by a phase angle that depends on the local strength of the turbulent mixing. It can be concluded that, while turbulent mixing affects both temperature and momentum transport, the analogy between heat transfer and momentum transfer is not applicable anymore, confirming the findings of (Ashton and Kennedy, 1972). This failure can be understood considering the presence of pressure anomalies induced by the ice ripples, which affect momentum transport but not temperature transport. An example of such pressure anomalies is provided in figure 6, which refers to a near-interface portion of the domain comprised between two consecutive ripples. A colormap is used to visualize the normalized pressure, $\overline{p}/\overline{p}_{max}$, in the water layer. Also plotted are the the iso-temperaure line $\mathcal{T} = 0.15\mathcal{T}_{max}$ (thick dotted line) and the iso-velocity line $u_x = 0.25u_{x,max}$ (thick dashed line). The two lines overlap near the first upstream ice crest on the left-end part of the plot, but separate downstream in the leeward portion of the ripple: Here, the iso-temperature line gets closer to the interface than the iso-velocity line. The separation is caused by a local adverse pressure gradient, which hinders momentum transfer from the bulk of the flow towards the ice-water interface, but does not affect heat transport. Further downstream, the pressure gradient becomes favorable but only momentum transfer is enhanced. As a result, the two iso-lines rejoin near the second crest. Overall, the effect of the pressure gradient is to shift the region of maximum shear stress downstream with respect to the region of maximum heat flux. We provide visual evidence of the shift in figure 7, where the time-averaged profiles of the normalized vertical heat flux, $\tilde{q}_z$, shear stress, $\tilde{\tau}_{xz}$, and melt rate, $\tilde{m} = -\dot{\xi}$, all evaluated over the same streamwise portion on the ice-water interface, are shown. The heat flux (solid orange line, panel a) exhibits an upstream shift relative to the shear stress (dashed blue line, panel a), leading to a larger phase angle measured relatively to the interface (dotted line): $\varphi_{q-\xi} > \varphi_{\tau-\xi}$. The same is found for the melt rate $\tilde{m}$ (solid purple line, panel b). The phase shifts of the heat flux and melt rate, computed from the cross-spectra $C_{\xi,q_z}$ and $C_{\xi,\dot{m}}$ (details on the computation are discussed in the Appendix A), are $\varphi_{q-\xi} \approx 0.59\pi$ and $\varphi_{\dot{m}-\xi} \approx 0.51\pi$, slightly above the instability threshold, while for the shear stress the value is $\varphi_{\tau-\xi} \approx 0.32\pi$, much lower than the instability threshold. The values of $\varphi_{q-\xi}$ and $\varphi_{\dot{m}-\xi}$, both close to $\pi/2$, indicate that the amplitude of the spanwise instability is nearly steady over time, implying that the main contribution to the ripple evolution is controlled by a characteristic migration velocity (Hanratty, 1981), much like the celerity of advancing desert or underwater dunes (Duran Vinent et al., 2019; Naqshband et al., 2014; Pähtz and Durán, 2020). Our findings indicate that this velocity is solely determined by melting and freezing, which regulate the morphodynamics of the ice-water interface, being considerably smaller than the typical velocity scales associated with the mean flow of the water stream and turbulence.

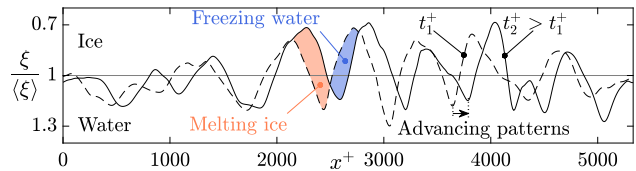

**Figure 4:** Morphodynamic evolution of the normalized ice thickness, $\xi/\langle\xi\rangle$, along the streamwise direction at two different times: $t_1^+$ (dashed line) and $t_2^+ > t_1^+$ (solid line).

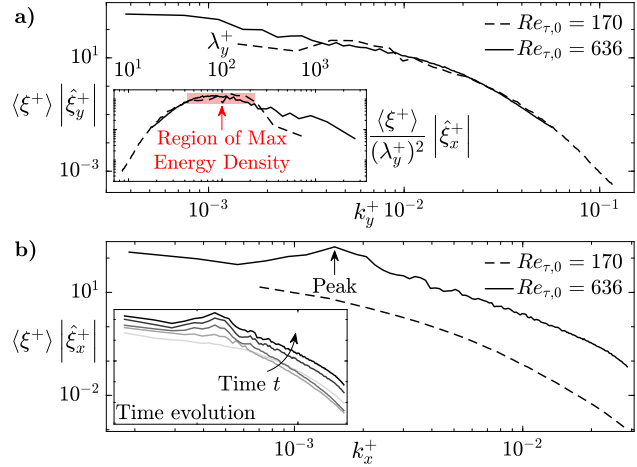

**Figure 5:** Ice thickness spectra along the spanwise direction ($|\hat{\xi}_y^+|$, panel a) and the streamwise direction ($|\hat{\xi}_x^+|$, panel b). Dashed lines: sub-critical $Re_{\tau,0} = 170$ case; solid lines: super-critical $Re_{\tau,0} = 636$ case. Values are computed taking the quadratic average of the spectra in time and space, along the streamwise direction for $|\hat{\xi}_y^+|$ and spanwise direction for $|\hat{\xi}_x^+|$. The inset in panel a) shows the pre-multiplied spanwise spectrum, $\langle \xi^+ \rangle |\hat{\xi}_y^+|/(\lambda_y^+)^2 = (k_y^+)^2 \langle \xi^+ \rangle |\hat{\xi}_y^+|$, as a function of the wavelength $\lambda_y^+$ and highlights the region of maximum energy density. The inset in panel b) shows the time evolution of the peak in the streamwise spectrum, $\langle \xi^+ \rangle |\hat{\xi}_x^+|$, over the entire span of the $Re_{\tau,0} = 636$ simulation.

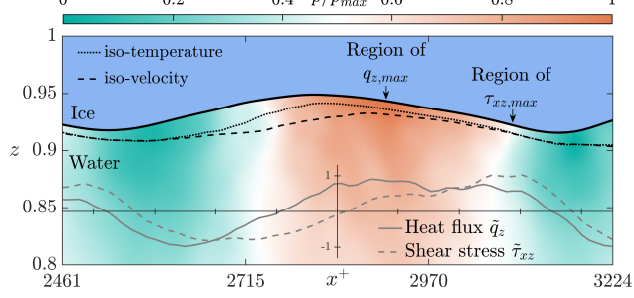

**Figure 6:** Iso-temperature ($\mathcal{T} = 0.15 \mathcal{T}_{max}$, thick dotted curve) and iso-velocity ($u_s = 0.25 u_{x,max}$, thick dashed curve) lines close to the ice-water surface (thick solid curve) between two consecutive crests. The colormap shows the normalized pressure, $\overline{p}/\overline{p}_{max}$, in the water. Also shown are the normalized vertical heat-flux, $\tilde{q}_z$ (thin solid curve), and the shear stress, $\tilde{\tau}_{xz}$ (thin dashed curve), at the interface: The region of maximum heat flux, $q_{z,max}$, is upstream of the region of maximum shear stress, $\tau_{xz,max}$. All quantities are time-averaged and refer to the same streamwise 2D slice of the domain.

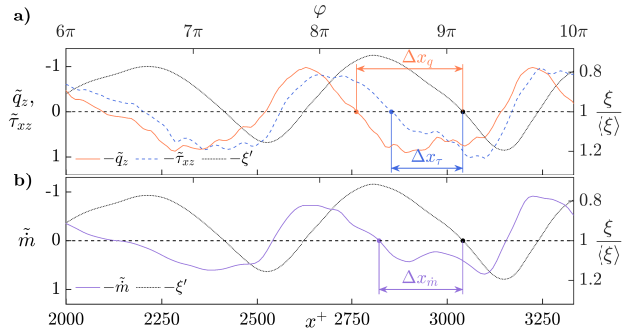

**Figure 7:** Time-averaged vertical heat flux, $\tilde{q}_z$ (solid orange line, panel a), shear stress, $\tilde{\tau}_{xz}$ (dashed blue line, panel a), and melt rate, $\tilde{m}$ (solid purple line, panel b), at the ice-water interface. For reference, the fluctuating component of the ice thickness, $\xi'$, is also shown (dotted line). The space lags $\Delta x_Q$ between the profile of $Q$ (with $Q$ representing melt rate, heat flux or shear stress) and the ice thickness, $\xi$, are reported. The angular phase $\varphi$ is given in the upper axis of panel a).

## 4. Conclusions

To gain predictive understanding of basal melting and freezing of ice, in this paper we use direct numerical simulation coupled with a phase field method and an immersed boundary method to study the morphodynamics of an ice layer over a turbulent stream of warm water. Through extremely accurate simulations, we provide a sound characterization of the melting/freezing phenomena that shape the ice-water interface, revealing new insights into the mechanisms that control the interface morphodynamics. At low water speeds, only streamwise undulations are observed and their formation is explained by the Reynolds analogy between heat and momentum transfer. However, for progressively higher velocities of the water stream, a threshold change of the interface morphology exists, which is controlled by the instability originating from the ice-water interactions. This instability leads to the formation of spanwise ripples that co-exist with the streamwise undulations, resulting in complex interface patterns that evolve along the water stream direction with a migration velocity much slower than the turbulence characteristic velocity. Our results demonstrate that the combined melting and freezing mechanisms triggered by turbulence and the morphodynamic instability cannot be explained by the Reynolds analogy, due to the occurrence of a phase shift between the local heat transfer and the local momentum transfer. We have been able to establish a causal relationship between the phase shift and the anomalies in pressure distribution and turbulent convection that are induced by the surface morphology. Considering the pivotal role played by melting and freezing in ice loss beneath ice shelves, we believe that our findings enhance the current understanding of ocean circulation within ice-shelf cavities and lay the groundwork for refining physics-based, geometry-dependent parameterizations of the melting process in large-scale ocean circulation models.

## Acknowledgments

We thank Dr. Francesco Zonta for the fruitful discussions in the early stages of the work. We acknowledge CINECA and VSC for HPC resources. D.P. gratefully acknowledges the generous funding from the national research programme PON Ricerca e Innovazione 2014-2020.

## A. Computation of the phase shift

The phase shift $\varphi_{q-\xi}$ between the local ice thickness $\xi$ and the local heat transfer $q_z$ has been computed by means of the spectrum $C_{\xi,q_z}$ of the cross-correlation $R_{\xi,q_z}$ along the streamwise direction $x$. The same computation has been performed to obtain the phase shift $\varphi_{\tau-\xi}$ between the local ice thickness and the local shear stress $\tau_{xz}$, as well as the phase shift $\varphi_{\dot{m}-\xi}$ between the local ice thickness and the local melt rate $\dot{m}$. Indicating with $Q$ any of the quantities cross-correlated with $\xi$, the following general definition can be applied:

$$R_{\xi,Q}(s) = \frac{1}{L_x L_y} \int_0^{L_y} \int_0^{L_x} \frac{\xi'(x-s,y)}{\langle \xi \rangle} \frac{Q'(x,y)}{\langle Q \rangle} \mathrm{d}x \mathrm{d}y,$$

where angular brackets $\langle \bullet \rangle$ indicate quantities that have been averaged in both $x$ and $y$, while the prime symbol indicates the fluctuating component of that quantity (e.g. $Q' = Q - \langle Q \rangle$). The cross-spectrum $C_{\xi,Q}$ is computed performing a Fourier transform:

$$C_{\xi,Q} = \mathcal{F}_x \left[ R_{\xi,Q} \right] = \frac{1}{L_x} \int_{-\infty}^{\infty} R_{\xi,Q}(s) \exp\left( -2\pi i s k_x \right) \mathrm{d}s.$$

Exploiting the properties of the Fourier transform, $C_{\xi,Q}$ can be also computed as follows:

$$C_{\xi,Q} = \frac{1}{L_y} \int_0^{L_y} \mathcal{F}_x \left[ \frac{\xi'}{\langle \xi \rangle} \right] \mathcal{F}_x \left[ \frac{Q'}{\langle Q \rangle} \right]^* \mathrm{d}y,$$

where the star symbol indicates the complex conjugate, which in its discrete form can be expressed as:

$$C_{\xi,Q}(k) = \frac{1}{L_y \langle \xi \rangle \langle Q \rangle} \sum_{j=i}^{N_y} \hat{\xi}'_x(k,j) \hat{Q}'_x(k,j)^*.$$

Here, the notation $\hat{\bullet}_x$ indicates the discrete Fourier transform in the $x$ direction. To compute the phase shift between $Q$ and $\xi$ (which is defined as positive if $Q$ precedes $\xi$ in the $x$ direction), the phase $\varphi$ of $C_{\xi,Q}$ must be evaluated at the characteristic wavenumber corresponding to the peak in the modulus, which is found to occur at $k_x = 8$ for each quantity $Q$ we analyzed.